\documentclass[pra,reprint,superscriptaddress]{revtex4-2}
\usepackage[utf8]{inputenc}
\usepackage[T1]{fontenc}
\usepackage{epsfig}
\usepackage{float}
\usepackage{mathptmx}
\usepackage{dcolumn}
\usepackage{bm}
\usepackage{siunitx}
\usepackage{graphicx}
\usepackage{times}
\usepackage{float}
\usepackage{amssymb}
\usepackage{amsmath}
\usepackage{amsfonts}
\usepackage{enumitem} 
\usepackage{afterpage}
\usepackage{babel}
\usepackage{setspace}
\usepackage[dvipsnames]{xcolor}
\usepackage{titlesec}
\usepackage{upgreek}
\usepackage{mathtools}
\usepackage[sort&compress]{natbib}

\setcitestyle{open={},close={},numbers,comma,super}

\usepackage[pagebackref=false]{hyperref}
\hypersetup{
    colorlinks  =   true,
    linkcolor   =   BlueViolet,
    citecolor   =   BlueViolet,
    filecolor   =   BlueViolet,
    urlcolor    =   BlueViolet}
\usepackage[capitalise,nameinlink]{cleveref}
\usepackage{hypcap}

\usepackage{titlesec}
\titleformat{\section}[hang]{\small\bfseries\sffamily}{\thesection.}{0.5em}{}
\titlespacing{\section}{0pc}{1.2pc}{0.3pc}
\titlespacing{\subsection}{0pc}{1pc}{0.2pc}

\makeatletter
\renewcommand*{\fnum@figure}{{\normalfont\bfseries \figurename~\thefigure}}
\renewcommand*{\@caption@fignum@sep}{\textbf{ | }}
\makeatother

\begin{document}

\title{Unveiling photon statistics with a 100-pixel photon-number-resolving detector\vspace{0.4cm}}

\author{Risheng Cheng}
\thanks{These authors contributed equally to this work.}
\affiliation{Department of Electrical Engineering, Yale University, New Haven, Connecticut 06511, USA.}

\author{Yiyu Zhou}
\thanks{These authors contributed equally to this work.}
\affiliation{Department of Electrical Engineering, Yale University, New Haven, Connecticut 06511, USA.}
\affiliation{Yale Quantum Institute, Yale University, New Haven, Connecticut 06511, USA.}

\author{Sihao Wang}
\author{Mohan Shen}
\author{Towsif Taher}
\affiliation{Department of Electrical Engineering, Yale University, New Haven, Connecticut 06511, USA.}

\author{Hong X. Tang}
\email[Corresponding author: ]{hong.tang@yale.edu}
\affiliation{Department of Electrical Engineering, Yale University, New Haven, Connecticut 06511, USA.}
\affiliation{Yale Quantum Institute, Yale University, New Haven, Connecticut 06511, USA.}

\date{\today}

\begin{abstract}
\vspace{0.15cm}
\noindent\textbf{Single-photon detectors are ubiquitous in quantum information science and quantum sensing. They are key enabling technologies for numerous scientific discoveries and fundamental tests of quantum optics. Photon-number-revolving detectors are the ultimate measurement tool of light. However, few detectors to date can provide high-fidelity photon number resolution at few-photon levels. Here, we demonstrate an on-chip detector that can resolve up to 100 photons by spatiotemporally multiplexing an array of superconducting nanowires along a single waveguide. 
The unparalleled photon number resolution paired with the high-speed response exclusively allows us to unveil the quantum photon statistics of a true thermal light source for the first time, which is realized by direct measurement of high-order correlation function $g^{(N)}$ with $N$ up to 15, observation of photon-subtraction-induced photon number enhancement, and quantum-limited state discrimination against a coherent light source. 
Our detector provides a viable route towards various important applications, including photonic quantum computation and quantum metrology.
}   

\end{abstract}
\maketitle


\noindent The observation and application of intriguing quantum optical effects have long been relying on high-performance photon-number-resolving (PNR) detectors \cite{hadfield2009single}. An ideal PNR detector should have unity detection efficiency, high speed, low timing jitter, negligible dark counts, and the ability of resolving photon numbers with a large dynamic range and high fidelity. Although transition edge sensors (TESs) \cite{nist_2003_pnr_tes,nist_2008_tes_95p,nist_2011_pnr_on-chip_tes,nist_2013_on-chip_tes,tes_100_photons} and microwave kinetic inductance detectors (MKIDs) \cite{gao_2017_pnr_mkid,day2003broadband} have demonstrated their inherent photon number resolvability as well as excellent low readout noise, their performances are mainly limited in terms of low count rate (typically below 1 MHz), large timing jitter (nanosecond level), and the need of ultra-low operation temperature ($\sim$100\,mK). On the other hand, superconducting nanowire single-photon detectors (SNSPDs)\cite{hadfield_2012_SNSPD_review} represent the cutting-edge technology with high efficiency (>98\%)\cite{nist_2019_98p_snspd,Chang2021}, gigahertz-level count rate \cite{simit_2019_16_pixel_detector}, picosecond-level timing jitter\cite{jpl_2020_3ps_jitter}, sub-hertz dark count noise\cite{NTT_2015_ultimate_darkcounts},  and an elevated operation temperatures (2--4\,K). However, the main drawback of SNSPDs is their limited photon number resolution. Although some modified readout schemes have been proposed to extend the PNR ability by either integrating a tapered impedance transformer\cite{berggren_2019_pnr_taper_snspd} or employing a wide-band cryogenic amplifier\cite{kim_2017_pnr_snspd}, the maximum resolvable photon number still remains at 3-4. 

Alternative approaches to construct a PNR-SNSPD include temporal-multiplexing\cite{hadfield_2013_time_multiplex_snspd} and spatial-multiplexing\cite{fiore_2008_pnr_snspd_parallel,cheng_2013_SND,fiore_2016_pnr_24,kit_2019_pnr_snspd_gaas, nist_2019_kilopixel_snspd}, where a sufficiently large pixel number is critical to achieve high PNR fidelity by minimizing the probability of more than one photons simultaneously firing the same detector pixel. The general limitation in the temporal-multiplexing scheme is the bulky setup consisting of multiple stages of long fibers to provide nanosecond-level delay, which is difficult to integrate on a single chip and thus has limited scalability. In comparison, the spatial-multiplexed array of SNSPDs have shown PNR of high dynamic range up to 24 photons\cite{fiore_2016_pnr_24}, while maintaining a simple readout scheme by series- or parallel-connecting all the detector pixels. Nevertheless, as the detected photon number increases, the resolution of the output voltage that is proportional to the photon number degrades significantly due to the limited signal-to-noise ratio and thus limit further scalability. In addition, both of these multiplexing schemes are lacking of the spatial resolution, which is highly desired in some quantum applications to resolve the spatial modes of the input photons. 

\begin{figure*}[!t]
\capstart
\centering
\includegraphics[width=0.9\linewidth]{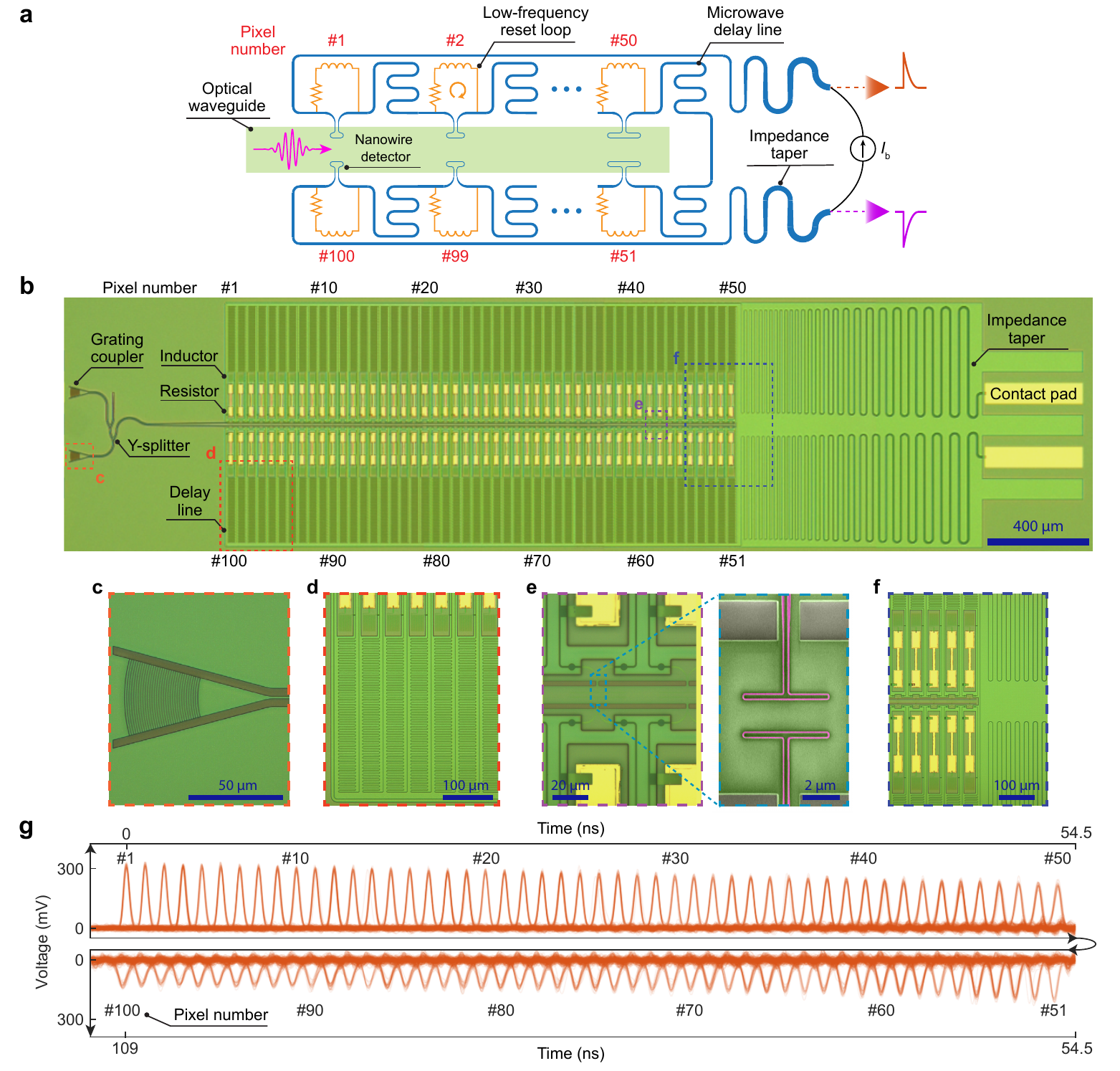}
\caption{\textbf{Device architecture and operation principle.}
\textbf{a}, Schematic illustration of the 100-pixel PNR detector structure based on the spatiotemporal-multiplexed SNSPD array. The high-frequency detector signals are read out through the bus line consisting of the series-connected nanowire detectors and delay line sections (blue color), while the bias currents of the fired detector pixels are recovered via the low-frequency reset loop formed by the on-chip inductor and resistor (orange color)  
\textbf{b}, Overview optical micrograph image of the device. The image is taken prior to finishing the device fabrication for better visibility (see Methods and Supplementary Section III). 
\textbf{c}, Expanded view of the grating coupler. 
\textbf{d}, Expanded view of the delay line sections located at the lower left corner of the device. 
\textbf{e}, Expanded optical micrograph and scanning-electron micrograph (SEM) view of the nanowire detectors with the leads passing through the waveguide bridges
\textbf{f}, Expanded view of the on-chip resistors, inductors, and part of the impedance tapers.   
\textbf{g}, Persistence oscilloscope traces of microwave pulses generated from each pixel of the detector. The presence/absence of the pulses combined with the arrival time of the pulse peaks can be used to provide both the total detected photon number and the position information of the fired pixels.}
\label{Fig:scheme}
\end{figure*}

Here, we demonstrate the design and experimental implementation of a waveguide-integrated and hybrid spatiotemporal-multiplexed SNSPD array, which features simultaneous photon number and position resolving capability, high scalability, and a simple readout scheme. In addition, the high timing resolution combined with the large-dynamic-range PNR capability of the detector allows us to analyze the photon number statistics of a true thermal light source. Thermal light has been playing a critical role in the development of quantum optics as it triggers the invention of intensity correlation $g^{(2)}$ measurement in the Hanbury Brown and Twiss experiment \cite{brown1956correlation}. Although thermal light has been extensively studied \cite{bennink2002two, zavatta2008subtracting, rafsanjani2017quantum, magana2019multiphoton}, most experiments use a pseudo-thermal light source that is artificially generated by passing a coherent laser light through a rotating ground glass \cite{goodman2015statistical, martienssen1964coherence}. This is because true thermal light sources typically have a broad bandwidth and thus a short coherence time, which places an extreme demand on the timing resolution of the single-photon detectors \cite{zhai2005two, liu2014lensless, tan2014measuring, deng2019quantum}. With the unprecedented photon number resolution as well as the high-speed responses of our detector, we are able to observe the photon statistics transition from a Bose-Einstein distribution to a Poisson distribution by tuning the pulse width of a true thermal light source. Our detector also permits a convenient direct measurement of the high-order correlation function $g^{(N)}$, with $N$ up to 15 in our experiment, which is only limited by the data acquisition rate of our test equipment, rather than the detector itself. We also implement the quantum-limited discrimination of a thermal and a coherent state approaching the Helstrom bound \cite{habif2021quantum} to showcase the potential of our detector in quantum metrology applications.


\begin{figure*}[!t]
\capstart
\centering
\includegraphics[width=0.9\linewidth]{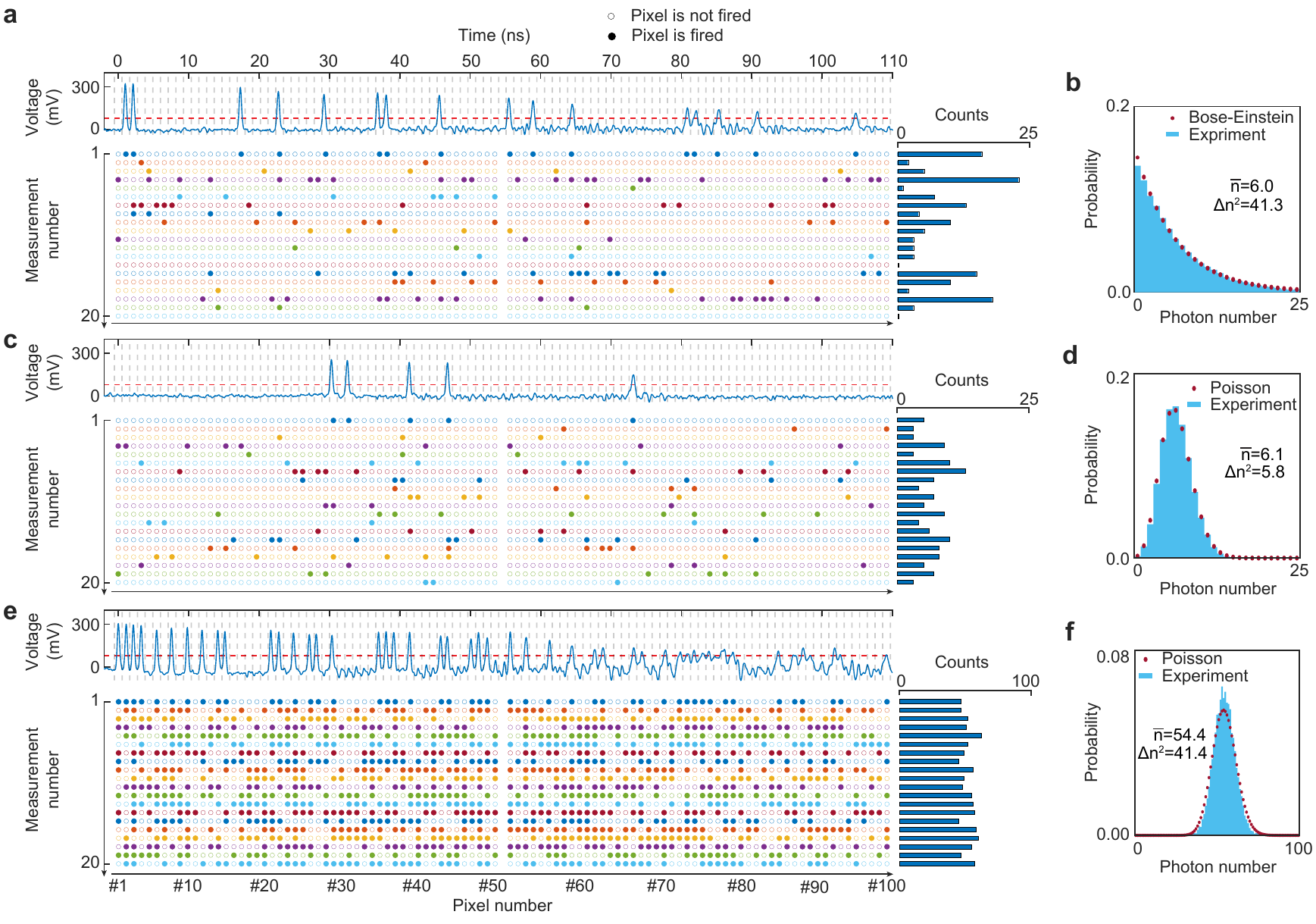}
\caption{\textbf{Photon statistics measurement.}
\textbf{a}, Example instance of the oscilloscope trace (upper panel), measurement sequences (lower panel), and detected counts of photon number (right panel).  By measuring the presence/absence of the pulses after an appropriate threshold (red dashed line) in each time slot, which corresponds to the individual pixel of the detector, the photon number as well as the position information on the fired pixels can be measured. 
\textbf{b}, Photon number probability distribution obtained by measuring the histogram of the detected photon numbers. The blue bars are the experimental data when EDFAs are used as the thermal light source, which agrees well with the red dots following the theoretical Bose-Eisenstein distribution with the mean photon number $\bar{n}=6.0$. 
\textbf{c}-\textbf{d}, Results for a laser as the coherent light source with $\bar{n}= 6.2$. \textbf{e}-\textbf{f}, Results for the coherent light source with $\bar{n}= 54.4$. 
}
\label{Fig:table}
\end{figure*}

\section*{\normalsize{}Results}
\noindent\textbf{Operation principle of detector.}
\cref{Fig:scheme}a illustrates the basic operation principle of the spatiotemporal-multiplexed SNSPD array. The incident photons are guided by the waveguide and spatially distributed to the nanowire detector pixels of incremental length integrated atop the waveguide. All the detector pixels are series connected with identical sections of nanosecond-level delay lines embedded between neighboring pixels. Each pixel is shunted by an on-chip inductor and resistor for the local reset of the bias current after firing. The fired pixel generates a pair of positive/negative microwave pulses that propagate along the opposite direction of the high-frequency bus line consisting of the detector pixels and the delay line sections (blue color). The signals are then read out in time sequence using the method of temporal-multiplexing (\cref{Fig:scheme}g). The impedance tapers are attached at both ends of the bus line to match the impedance between the delay lines and the reading electronics, which is critical for minimizing reflections of the readout signals, and thus preserving clear and fast rising/falling edges\cite{berggren_2017_single_photon_imager,cheng_2019_broadband_spectrometer}. In addition, the presence of the resistively shunted inductors prevents the high-frequency signals entering the low-frequency reset loop (orange color), rendering the readout and reset of the individual detector pixels separated.    

One advantage of this device design is that the nanowire detectors, on-chip inductors, microwave delay lines as well as the impedance tapers could be all implemented in a single ultra-thin niobium nitride (NbN) superconducting layer (\cref{Fig:scheme}b-f), thus greatly simplifying the fabrication process. Moreover, a compact design of ultra-slow microwave transmission line based on high-kinetic-inductance NbN nanowires and high-\textit{k} dielectric layer (see Methods) is employed here instead of bulky optical delay lines for the temporal-multiplexed readout, and thus a high scalability of the detector pixel number is expected. More importantly, this hybrid spatiotemporal-multiplexing scheme enables position resolving capability; that is, in addition to the information of photon number, the readout scheme could also provide position information on the fired detector pixels. This distinct property compared to other conventional PNR detectors makes our detector a highly versatile platform to perform various quantum optics experiments on a chip.     

\begin{figure*}[!t]
\capstart
\centering
\includegraphics[width=\linewidth]{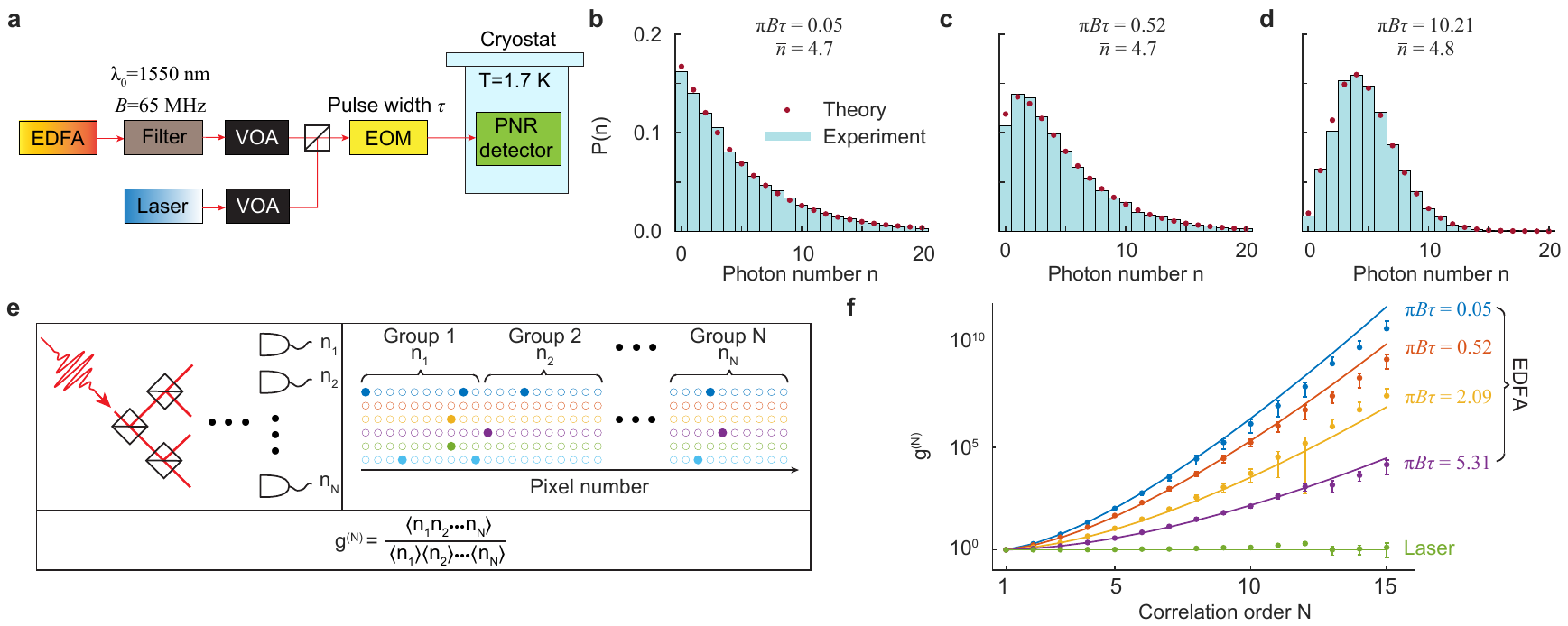}
\caption{\textbf{Photon statistics and high-order correlation measurement}.
\textbf{a}, Simplified schematic of the experimental setup. VOA, variable optical attenuator. EOM, electro-optic modulator. PNR detector, photon-number-resolving detector.
\textbf{b}-\textbf{d}, Photon number probability distribution measured with spectrally filtered EDFAs used as the thermal light source. We change the EOM pulse width $\tau$ to tune the time-bandwidth product $\pi B \tau$, which is labeled at the top of each panel.
\textbf{e}, Comparison between the conventional method and our implementation to measure the high-order correlation function $g^{(N)}$.  In our implementation, we split the 100-pixel detector into $N$ groups and use the separate sum of photon counts in the $N$ groups for coincidence measurement. This method is equivalent to the conventional method where the input light beam is split by a beam-splitter tree into $N$ paths, each of which is detected by a PNR-detector.   
\textbf{f}, Measured $g^{(N)}$ with $N$ up to 15 (dots) in comparison with the theoretical calculation (solid curves). The type and parameters used for the light source are labeled beside each corresponding curves. In particular, we experimentally measure $g^{(2)}=1.961 \pm 0.003$ for $\pi B \tau=0.05$.
}
\label{Fig:transition}
\end{figure*}
\noindent\textbf{Photon statistics and correlation measurement.}
We first demonstrate the potential of our detector in quantum optics applications by measuring the photon statistics of thermal and coherent states of light. A fast-speed oscilloscope is used to acquire microwave pulse signals generated by the detector and amplified by low-noise amplifiers (see Supplementary Section I for more details on the measurement setup). One instance of the oscilloscope trace produced by detecting a thermal light source from erbium-doped fiber amplifiers (EDFAs) is shown in \cref{Fig:table}a. The presence/absence of the pulses in each time slot, which corresponds to the individual pixel of the detector, can be used to provide the total detected photon number as well as the position information of the fired pixels. The lower panel of \cref{Fig:table}a shows 20 measurement sequences as example, and the corresponding total count of photon number in each measurement is attached to the end of each sequence. By collecting a sufficient number of measurements, the histogram of the detected photon numbers can be constructed to get the photon number probability distribution (see \cref{Fig:table}b), which follows the Bose-Einstein distribution in excellent agreement. The statistics measurement of a coherent laser source with a similar mean photon number ($\bar{n}=6.2$) is shown in \cref{Fig:table}c,d and a much higher number ($\bar{n}=54.4$) in \cref{Fig:table}e,f to showcase the large-dynamic-range PNR capability of our detector. Both traces from the coherent source follow the Poisson distribution, and the counts of photon number fluctuate much less strongly than that of the thermal state due to the lack of photon bunching (also see Supplementary Movies). It is also worth noting that as the total detected photon number increases, the peak-to-background ratio gets deteriorated, particularly for the signals from the latter half part of the pixels. This is due to the accumulated low-frequency background ripples from the fired pixel signals in front, which is also captured in our simulation (Supplementary Section II). To mitigate this problem, we can use both output channels of our detector and separately collect the signals for \#1\,-\,\#50 and \#51\,-\,\#100 pixels. In the future upgrade, a higher-permittivity dielectric layer, such as SrTiO$_3$\cite{Santavicca2021}, can be employed to greatly reduce the propagation speed of the microwave signals and thus increase the delay time between neighboring pulse signals to help differentiate them. In addition, more advanced waveform processing schemes\cite{berggren_2018_scalable_detector} could be developed in the future, instead of simple pulse peaks  counting, to better extract the detection information.       

\begin{figure*}[!t]
\capstart
\centering
\includegraphics[width=\linewidth]{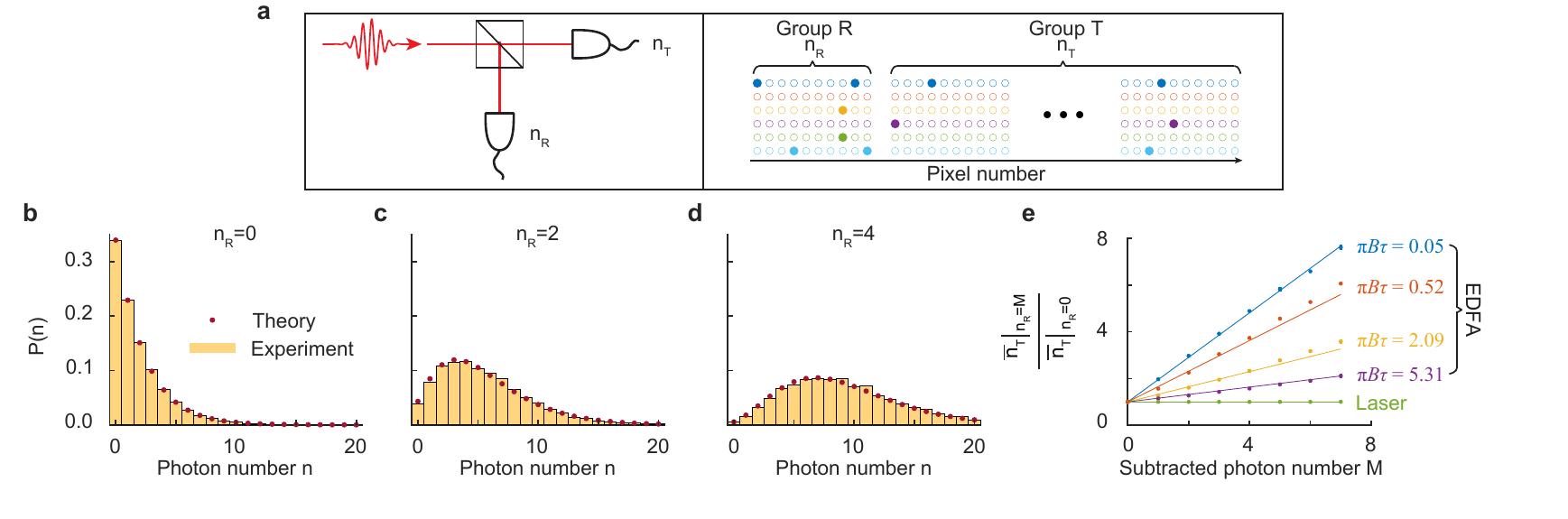}
\caption{\textbf{Photon subtraction experiment.}
\textbf{a}, Comparison between the conventional method and our implementation to measure the photon number statistics for a photon-subtracted thermal light. The conventional method employs a beam-splitter and measure the photon number $n_R$ and $n_T$ at the reflection and transmission port using two PNR-detectors, respectively. In our implementation, we split the 100-pixel detector into two groups and measure the photon number probability of group T in the event of $n_R=M$.
\textbf{b}-\textbf{d}, Photon number probability distribution with various subtracted photon number $n_R=0, 2, 4$, respectively. The spectrally filtered EDFAs with $\pi B \tau=0.05$ are used as the thermal light source.
\textbf{e}, Photon number enhancement from measurement (dots) and theoretical calculation (solid curves). The source type and parameters used for the light source are labeled beside each corresponding curves.}
\label{Fig:subtraction}
\end{figure*}

We next use our detector to investigate the effect of pulse width on the quantum statistics of a true thermal light source and compare with that of a coherent counterpart. A simplified schematic of the experimental setup is shown in \cref{Fig:transition}a (see Supplementary Section I for a complete schematic). The spectrally filtered amplified spontaneous emission from EDFAs and the narrow-bandwidth coherent-state light from a continuous-wave (CW) laser are separately used as the light sources with the variable optical attenuators (VOAs) tuning the light intensity and thus the mean photon number $\bar{n}$. Both of the sources are followed by the temporal gating through the electro-optic modulator (EOM) to make them pulsed sources. The photon number probability distribution of a thermal light measured by a PNR detector can be described by a negative binomial distribution as \cite{karp2013optical,  agarwal1992negative}

\begin{equation}
\begin{split}
P_{\text{NB}}(n; \pi B \tau, \bar{n}) &= \binom{n+\pi B \tau}{n} \left( \frac{\pi B \tau +1}{\bar{n} + \pi B \tau + 1} \right)^{\pi B \tau +1} \\
&\times \left( \frac{\bar{n}}{\bar{n} + \pi B \tau + 1} \right)^n ,
\end{split}
\label{eq:NB}
\end{equation}

where $B$ is the full width at half maximum (FWHM) bandwidth of the Lorentzian optical spectral filter \cite{mandel1962measures}, and $\tau$ the pulse width. When the time-bandwidth product $\pi B \tau \ll 1$, \cref{eq:NB} can be reduced to the Bose-Einstein distribution $P_{\text{BE}}(n; \bar{n}) = \frac{1}{ \bar{n} +1 } \left( \frac{\bar{n}}{ \bar{n} +1 } \right)^n $. On the other hand, when $\pi B \tau \gg 1$, it becomes the Poisson distribution $P_{\text{P}}(n; \bar{n}) = \bar{n} ^n e^{-\bar{n}} / n!$. The spectral filter used in our experiment has a bandwidth of $B=65$ MHz, and the pulse width is tuned by the EOM from 0.24\,ns to 100\,ns. The photon number probability distributions for $\pi B \tau  = 0.05, 0.52, 10.21$ are shown in \cref{Fig:transition}b-d, respectively. It can be clearly seen that the photon statistics gradually transits from the Bose-Einstein distribution to the Poisson distribution, which agrees well with the theoretical calculation by \cref{eq:NB}. 

With the capability of the position resolving, our detector also allows a convenient and direct measurement of the high-order correlation function $g^{(N)} = \langle n_1 n_2 \cdots n_N \rangle / \langle n_1 \rangle \langle n_2 \rangle \cdots \langle n_N \rangle$, where $\langle \cdot \rangle$ indicates the ensemble average. The previous conventional measurement of $g^{(N)}$ is typically realized by splitting an optical beam into $N$ parts and subsequently using $N$ single-photon detectors for coincidence counting. By contrast, we can flexibly split 100 pixels of our detector into $N$ groups without modifying the experiment setup and count the sum photon number in each group for coincidence measurement, as illustrated in \cref{Fig:transition}e. The theoretical value of $g^{(N)}$ for the EDFA thermal light source can be calculated as \cite{zhai2013photon}
\begin{equation}
g^{(N)} = \frac{\Gamma(\pi B \tau +N+1)}{\Gamma(\pi B \tau + 1) (\pi B \tau + 1)^N},
\end{equation}
where $\Gamma( \cdot )$ is the gamma function. In the limit of $\pi B \tau \ll 1$, we have $g^{(N)} \rightarrow N!$, which agrees with the result of the Bose-Einstein distribution \cite{stevens2010high}. At the other extreme $\pi B \tau \gg 1$, it follows that $g^{(N)} \rightarrow 1$, the well-known results of the Poisson distribution. The experimental results for $g^{(N)}$ with $N$ up to 15 are shown in \cref{Fig:transition}f, which agree well with the theoretical prediction. When $\pi B \tau$ decreases, we can see that the value of $g^{(N)}$ increases significantly. By contrast, the value of $g^{(N)}$ remains unity when a coherent laser is used as the light source. It should be noted that the measurement of $g^{(N)}$ with higher $N>15$ is limited by the slow data  acquisition rate of the oscilloscope ($\sim$30\,Hz) used in our experiment, rather than the detector itself, since an exponentially larger sample number is needed to minimize the variance of $g^{(N)}$ as $N$ increases. A dedicated GHz-level fast pulse counter could be employed in the future experiment to further unlock the performance of our detector.      

\noindent\textbf{Photon subtraction experiment.}
We then further investigate the capability of our detector by performing photon subtraction experiment. The conventional photon subtraction entails using a beam-splitter to probabilistically subtract a certain number of photons at the reflection port as shown in \cref{Fig:subtraction}a. In our experiment, we split the 100-pixel detector into two groups, i.e., group R and group T, and a splitting ratio $R:T=20\%:80\%$ is achieved by adjusting the number of pixels in each group. Due to the bunching effect of thermal light, the more photons are subtracted at the reflection port, the higher mean photon number will be detected at the transmission port \cite{parigi2007probing, barnett2018statistics}. This effect can be potentially useful for applications, such as quantum-enhanced interferometry \cite{rafsanjani2017quantum} and quantum-state engineering \cite{magana2019multiphoton}. Assuming that the input photon number probability distribution can be described by a negative binomial distribution $P_{\text{NB}}^{\text{in}}(n;\pi B \tau, \bar{n}_{\text{in}})$ with a mean photon number of $\bar{n}_{\text{in}}$, the photon number probability distribution at the transmission port can be described by a second negative binomial distribution $P_{\text{nb}}^{\text{T}}(n; \pi B \tau + n_{\text{R}}, \bar{n}_{\text{T}})$, whose mean photon number $\bar{n}_{\text{T}}$ becomes
\begin{equation}
\bar{n}_{\text{T}} = \frac{(1-R)(n_\text{R} + \pi B \tau  + 1)}{R \bar{n}_{\text{in}} + \pi B \tau + 1} \bar{n}_{\text{in}} ,
\end{equation}
where $n_{\text{R}}$ is the subtracted photon number at the reflection port. When $R \rightarrow 0$, $\pi B \tau \ll 1$, and $n_{\text{R}}=1$, we have $\bar{n}_{\text{T}}=2 \bar{n}_{\text{in}}$, which is the well-known result that the mean photon number of a thermal state at the transmission port will be doubled when a single photon is annihilated at the reflection port of a low-reflection beam-splitter \cite{zavatta2008subtracting}. It is also worth noting that the time-bandwidth product of the negative binomial distribution at the transmission port increases from $\pi B \tau$ to $\pi B \tau + n_{\text{R}}$. Therefore, if the input beam follows the Bose-Einstein statistics with $\pi B \tau 
\ll 1$, the photon number distribution at the transmission port will transit to Poisson-like statistics when the annihilated photon number $n_{\text{R}}$ is large. These theoretical predictions agree well with our experimental data as shown in \cref{Fig:subtraction}b-d, where $n_{\text{R}} = 0, 2, 4$ respectively. For these measurement, the spectrally filtered EDFA is used as the light source with $\pi B \tau = 0.05$. 

We next investigate the photon number enhancement effect of photon subtraction, and we use 
\begin{equation}
\frac{\bar{n}_{\text{T}} | _{n_{\text{R}}=M}}{\bar{n}_{\text{T}} | _{n_{\text{R}}=0}} = \frac{\pi B \tau + M + 1}{\pi B \tau + 1}
\end{equation}
to quantify the photon number enhancement, where $M$ is the subtracted photon number at the reflection port. The experimental results of photon number enhancement are shown in \cref{Fig:subtraction}e, which illustrates that the photon number enhancement increases linearly as $M$ increases. In contrast, there is no enhancement for the coherent light source from the laser, indicating the photon states at the two output ports of the beam-splitter are uncorrelated for a coherent state input\cite{zavatta2008subtracting}.

\noindent\textbf{Quantum-limited discrimination of photon states.}
As another application of our detector, we perform quantum-limited discrimination between a thermal state and a coherent state of light \cite{habif2021quantum}. Here, the mean photon number $\bar{n}$ of the light source is assumed to be known, but its quantum state, either a thermal state or a coherent state, remains to be discriminated. The minimum possible discrimination error is known as the Helstrom bound \cite{helstrom1969quantum}, which can be computed as $P^{\text{H}}_{\text{err}}(\bar{n}) = 0.5 - 0.25 || \rho_{\text{th}}(\bar{n}) - \rho_{\text{coh}} (\bar{n}) ||_1 $, where $\rho_{\text{th}}(\bar{n})$ is the density matrix of a thermal state, $\rho_{\text{coh}}(\bar{n})= | \alpha=\sqrt{\bar{n}} \rangle \langle \alpha |$ is the density matrix of a coherent state, and $||\cdot ||_1$ is the trace distance.

\begin{figure}[t]
\capstart
\centering
\includegraphics[width=\linewidth]{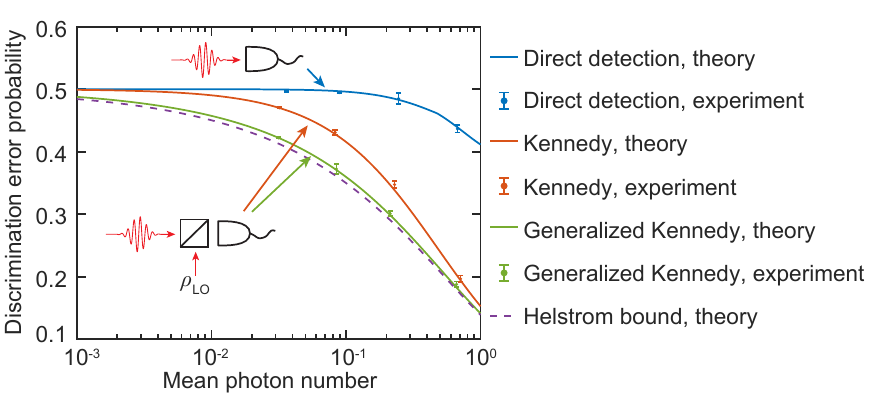}
\caption{\textbf{Quantum-limited discrimination of coherent and thermal states}.
The theoretical values of the discrimination error probability are represented by solid lines, while the experimental results are represented by dots and standard deviation error bars. The insets show  simplified schematics for the corresponding measurement methods.}
\label{Fig:helstrom}
\end{figure}

The most straightforward method for discrimination is the direct detection, which entails using a PNR detector to measure the photon number emitted by the source. When $n$ photons are measured, its quantum state is discriminated to be a thermal state if $P_{\text{BE}}(n; \bar{n}) > P_{\text{P}}(n; \bar{n})$. On the other hand, if $P_{\text{BE}}(n; \bar{n}) \leq P_{\text{P}}(n; \bar{n})$, the quantum state is discriminated to be a coherent state. The error probability of discrimination can thus be calculated as $P_{\text{direct}}^{\text{err}}(\bar{n}) = \sum_{n} \min (P_{\text{BE}}(n;\bar{n}), P_{\text{P}}(n;\bar{n})) / 2$. However, this method cannot approach the Helstrom bound. The Kennedy receiver can outperform the direct detection by using homodyne detection. A high-transmission beam-splitter ($T \approx 1$) is used to reduce the loss for the signal to be discriminated. The Kennedy detection entails using a local oscillator with a mean photon number of $\bar{n}T/(1-T)$ and a relative $\pi$ phase shift with respect to the coherent state light source, whose density matrix can be expressed as $\rho_{\text{LO}}= | -\alpha_{\text{LO}} \rangle \langle -\alpha_{\text{LO}} |$ with $\alpha_{\text{LO}} = \sqrt{\bar{n}}$. If the source is a coherent state, the transmission port will present a vacuum state due to the destructive interference. If the source is a thermal state, the state at the transmission port possesses a Laguerre statistics \cite{karp2013optical,habif2021quantum} $P_{\text{L}}(n) = \frac{(\bar{n}T)^n}{(1+\bar{n}T)^{n+1}} \exp (-\frac{\bar{n}T}{1+\bar{n}T}) L_n (-\frac{1}{1+\bar{n}T})$, where $L_n(\cdot)$ is the Laguerre polynomial of order $n$. Therefore, the discrimination error probability is $P_{\text{K}}^{\text{err}}(\bar{n}) =  P_{\text{L}}(0) / 2$. The Kennedy detection, however, can be further generalized to approach the Helstrom bound by changing $\alpha_{\text{LO}}$ to $\alpha_{\text{LO}}=\sqrt{\bar{n} + \Delta n}$, where the value of $\Delta n$ depends on $\bar{n}$ and can be computed by numerical optimization \cite{habif2021quantum}. In this case, if the source is a coherent state, the detected statistics becomes a Poisson distribution $P_{\text{P}}(n;\Delta n) $ with a mean photon number of $\Delta n$. If the source is a thermal state, the detected statistics becomes a Laguerre distribution $P_{\text{L}}(n,\Delta n) = \frac{(\bar{n}T)^n}{(1+\bar{n}T)^{n+1}} \exp(-\frac{(\bar{n}+\Delta n)T}{1+\bar{n}T}) L_n (-\frac{\bar{n}+\Delta n}{\bar{n}(1+\bar{n}T)} )$, and thus the discrimination error probability is $P_{\text{GK}}^{\text{err}}(\bar{n}) = \sum_{n} \min (P_{\text{P}}(n;\Delta n), P_{\text{L}}(n, \Delta n)) / 2$. The experimental results measured by the different methods are shown in \cref{Fig:helstrom}, which demonstrate that the generalized Kennedy detection closely approaches the Helstrom bound and is the best discrimination method that has been reported \cite{habif2021quantum}.

\section*{\normalsize{}Discussions}

\noindent In this Article, we have demonstrated a waveguide-integrated and large-dynamic-range PNR detector that employs a hybrid spatiotemporal-multiplexing scheme to read out a 100-pixel SNSPD array. In addition to inheriting all aspects of the excellent performances of single-pixel SNSPDs, this scheme also enables readout of the position information on the fired detector pixels. This distinct functionality makes our detector a versatile platform to perform various quantum optics
experiments on a chip at an unprecedented level, such as the photon statistics measurement, the direct measurement of high-order correlation function $g^{(N)}$ with $N$ up to 15, the photon subtraction experiment with the subtracted photon number up to 4, and the quantum-limited photon state discrimination. In particular, these measurements present a comprehensive characterization of the quantum statistical properties of a true thermal light source for the first time, which is exclusively enabled by the simultaneous availability of high photon number resolution and fast detector response. We believe that our novel detector technology can boost a variety of applications in the near future, such as large-scale Boson sampling \cite{tillmann2013experimental}, photonic quantum computing \cite{kok2007linear} and quantum metrology \cite{matthews2016towards}.


\def\bibsection{\section*{References}}
\bibliographystyle{Risheng}
\bibliography{My_reference}

\section*{{\normalsize{}Acknowledgments}}

{\noindent\footnotesize{}This work is funded by the National Science Foundation (NSF) through ERC Center for Quantum Networks (CQN) grant EEC-1941583. We acknowledge early funding support for this project from DARPA DETECT program through an ARO grant (No. W911NF-16-2-0151) and NSF EFRI grant (EFMA-1640959).  The authors would like to thank Dr. Yong Sun, Sean Rinehart, Kelly Woods, and Dr. Michael Rooks for their assistance provided in the device fabrication. The fabrication of the devices was done at the Yale School of Engineering \& Applied Science (SEAS) Cleanroom and the Yale Institute for Nanoscience and Quantum Engineering (YINQE).}{\footnotesize\par}

\section*{{\normalsize{}Author contributions }}

{\noindent\footnotesize{}R.C., Y.Z., and H.X.T. conceived the idea and experiment; R.C. designed and fabricated the devices; R.C., Y.Z., S.W., M.S., and T.T. performed the measurements; R.C. and Y.Z. analyzed the data. R.C., Y.Z. and H.X.T. wrote the manuscript with inputs from all authors. H.X.T. supervised the project.}{\footnotesize\par}

\section*{{\normalsize{}Competing interests }}

{\noindent\footnotesize{}The authors declare no competing interests. }{\footnotesize\par}

\section*{{\normalsize{}Additional information}}
{\noindent\footnotesize{}\textbf{Supplementary information} is available in the online version of the paper.}

{\noindent\footnotesize{}\textbf{Correspondence and requests for materials} should be addressed to H.X.T.}
{\footnotesize\par}

\afterpage{}
\clearpage
\section*{{\normalsize{}Methods}}
\noindent\footnotesize{}\textbf{Device fabrication.} 
The fabrication of the device involves six steps of high-resolution ($100\,\mathrm{kV}$) e-beam lithography (EBL) and following etching or deposition/lift-off process. (1) Alignment markers are pre-fabricated on top of a 4-inch 330$\,\mathrm{nm}$-thick LPCVD-Si$_3$N$_4$ / \SI{3.3}{\micro\meter}-thick thermal oxide / Si wafer by standard double-layer polymethyl methacrylate (PMMA) patterning process, e-beam evaporation of 10\,nm Cr\,/\,100\,nm Au layers, and following lift-off in acetone. (2) High-uniformity 8\,nm-thick NbN superconducting thin film is deposited by atomic layer deposition (ALD)\cite{cheng_2019_snspd_ald} afterwards and followed by the e-beam patterning of the negative-tone 6\% hydrogen silsesquioxane (HSQ) resist. It is worth noting that the spinning and patterning of HSQ resist right after the fresh NbN film deposition is critical in maintaining excellent adhesion of the resist. (3) In the third EBL step, the electrode pads and contact pads for the on-chip resistors are defined by another double-layer PMMA patterning, e-beam evaporation of 5\,nm Cr\,/\,25\,nm Au, and following lift-off process. Later, the HSQ pattern is transferred to the NbN layer in a timed reactive-ion etching (RIE) step employing tetrafluoromethane (CF\textsubscript{4}) chemistry, which defines the nanowire detector pixels, delay lines, on-chip inductors, and impedance tapers.  (4) In the fourth EBL step, we pattern the positive-tone AR-P 6200 (CSAR 62) resist for the photonic microstructures, including gratings couplers, Y-splitters, and waveguides. The patterns are then transferred to the Si$_3$N\textsubscript{4} film via carefully timed RIE in fluoroform (CHF\textsubscript{3}). The remaining resist is removed by gentle oxygen plasma and hot N-methylpyrrolidone (NMP). (5) In the fifth EBL step, we define the on-chip resistors by patterning double-layer PMMA and e-beam evaporation of 80\,nm Cr\,/\,5\,nm Au with lift-off process. (6) Lastly, the 80\,nm-thick alumina (AlO\textit{\textsubscript{x}}) dielectric spacing layer and 100\,nm-thick aluminum (Al) top ground layer are fabricated by double-layer PMMA patterning, e-beam evaporation, and lift-off for making the low-speed microwave transmission line. The remaining HSQ mask on top of the NbN delay line area is removed by 1:100 diluted hydrofluoric (HF) acid after opening the PMMA window (lighter area in \cref{Fig:scheme}b) and before the deposition of AlO\textit{\textsubscript{x}}/Al layers. The resulting devices is shown in \cref{Fig:scheme}b-f, and additional device images as well as the fabrication yield are presented in Supplementary Section III.

\noindent\footnotesize{}\textbf{Device design.} 
As shown in \cref{Fig:scheme}b, the photons are launched into the Si$_3$N$_4$ waveguide from a fiber through the input grating coupler. The waveguide is then split into two branches, one of which is attached to the output grating coupler for power calibration and alignment, while the other one is adiabatically tapered from 1.2\,$\upmu$m width to 8\,$\upmu$m and then connected to a series of 100 NbN nanowire detector pixels placed atop the waveguide. All the detector pixels are made of hairpin-shaped 80\,nm-wide NbN nanowires that guarantee saturated internal efficiency at 1550\,nm telecommunication wavelength (see Supplementary Section V). Each of the nanowire is shunted by a 5\,$\upmu$m-wide and 50\,$\upmu$m-long Cr/Au on-chip resistor (corresponding to 60\,$\Upomega$ resistance at 1.7\,K) and an inductor made of a 400\,nm-wide NbN meander wire (corresponding to 200\,nH inductance). All the nanowires are series-connected with long meandered 800\,nm-wide NbN delay lines embedded in between, as visualized in \cref{Fig:scheme}b. Both ends of the nanowire-detector-delay-line chain are connected to the impedance tapers consisting of meandered NbN wires with gradually varying widths from 800\,nm to 9.2\,$\upmu$m, which follow the Klopfenstein design to match the impedance between the delay lines (hundreds of ohms) and the external reading electronics (50\,$\Upomega$). The total length of each impedance taper is $24\,\mathrm{mm}$, which allows <50\,MHz lower cut-off frequency. The long impedance tapers here play an important role in preserving the fast-rising edges and clear shapes of the photon-excited detector pulses by suppressing multiple reflections between the device and external readout electronics. At the ends of the impedance tapers, the microstrip lines are converted to coplanar waveguides (CPWs) to match the modes of ground-signal-ground (GSG) RF probes.

By engineering the length and the distance of the nanowire from the edge of the waveguide, we could tune the photon absorption efficiency of each nanowire and thus ensure the input photons are evenly distributed over the detector pixels. In the etched clearance area for the waveguide, a 1.3\,$\upmu$m-wide bridges are designed for the connection of each nanowire detector to the delay line sections that are positioned outside the waveguide area (\cref{Fig:scheme}e). The 3D finite-difference time-domain (FDTD) simulation study shows that the total scattering loss induced by 100 of such bridges is lower than 0.3\,dB. We further confirm the results by investigating the quality factors of Si$_3$N$_4$ racetrack resonators with varying number of bridges embedded in the straight waveguide sections.  

The parts of NbN delay lines and the impedance tapers are capped by a high-dielectric-constant (high-\textit{k}) AlO\textit{\textsubscript{x}} layer and top metal ground made of Al layer (lighter area in \cref{Fig:scheme}b), which helps significantly slow down the propagation speed of the photon-excited microwave signals, thereby enabling a very compact device design and also reducing the cross-talk between the transmission lines. Due to the high kinetic inductance of the superconducting wires made of NbN thin film and the high capacitance introduced by the thin high-$k$ layer, the propagation speed could be achieved as low as 1.07\% of $c$ (vacuum speed of light), despite the significantly wide delay lines compared to the nanowire detector. Each delay line section is designed to be 3.4\,mm long and introduces a 1.09\,ns propagation time delay between neighboring detector pixels (\cref{Fig:scheme}g).

\noindent\footnotesize{}\textbf{Device characterization.} 
The sample chip containing tens of detector devices is mounted on a 3-axis stack of stages (Attocube) inside a low-vibration closed-cycle refrigerator (Cryomech) and cooled down to $1.7\,\mathrm{K}$ base temperature. We drive the stages to move the sample chip and make the electrode contact with a multi-channel RF probe, which is used for DC-biasing the detector device and reading out the photon-excited microwave pulses. In the meantime, the grating couplers are aligned to the single-mode fiber array by optimizing the optical transmission through the device (see \cref{Fig:scheme} and Supplementary Section V). The photons are coupled into the waveguide from one of the fiber through the input grating coupler, while parts of them are output from the other calibration grating coupler, collected with another optical fiber in the same fiber array, and finally detected with a calibrated room-temperature photodetector.
This cryogenic optical setup not only guarantees multiple optical and RF channels input/output, but also allows for measurements on multiple devices in a single round of cool-down. 

The pulsed coherent light source is generated by gating a tunable continuous-wave (CW) laser (New Focus 6427 Vidia-Discrete) with the cascaded electro-optic modulators (EOMs) and acoustic-optic modulator (AOM), which guarantees sub-ns narrow pulse width combined with >100\,dB extinction ratio to efficiently suppress the background photons. The thermal light source is generated by employing the same gating method but with the EDFA/narrow-bandwidth spectral filter chain to replace the CW laser part. A more detailed description on the measurement setup can be found in Supplementary Section I.

\section*{{\normalsize{}Data availability}}
\noindent\footnotesize{} The data that support the plots within this paper and other findings of this study are available from the corresponding author upon reasonable request.

\end{document}